\documentclass[aps,pra,reprint,superscriptaddress,twocolumn,showkeys,amsmath,amssymb,longbibliography]{revtex4-1}
\usepackage[english]{babel}
\usepackage{amsmath,amssymb,bbm,mathrsfs,bm,braket,color,graphicx,comment,amsfonts,dsfont}
\usepackage[colorlinks,citecolor=blue,urlcolor=blue]{hyperref}
\usepackage{xcolor}
\usepackage[mathscr]{euscript}

\newcommand{\pd}{{\phantom\dag}}
\begin{document}

\title{Temperature enhancement of thermal Hall conductance quantization}

\author{I. C. Fulga}
\affiliation{IFW Dresden and W{\"u}rzburg-Dresden Cluster of Excellence ct.qmat, Helmholtzstrasse 20, 01069 Dresden, Germany}

\author{Yuval Oreg}
\affiliation{Department of Condensed Matter Physics, Weizmann Institute of Science, Rehovot 76100, Israel}

\author{Alexander D. Mirlin}
\affiliation{Institute for Quantum Materials and Technologies, Karlsruhe Institute of Technology, 76021 Karlsruhe, Germany}
\affiliation{Institut f\"{u}r Theorie der Kondensierten Materie, Karlsruhe Institute of Technology, 76128 Karlsruhe, Germany}
\affiliation{L.\,D.~Landau Institute for Theoretical Physics RAS, 119334 Moscow, Russia}

\author{Ady Stern}
\affiliation{Department of Condensed Matter Physics, Weizmann Institute of Science, Rehovot 76100, Israel}

\author{David F. Mross}
\affiliation{Department of Condensed Matter Physics, Weizmann Institute of Science, Rehovot 76100, Israel}

\date{\today}
\begin{abstract}
The quest for non-Abelian quasiparticles has inspired decades of experimental and theoretical efforts, where the scarcity of direct probes poses a key challenge. 
Among their clearest signatures is a thermal Hall conductance with quantized half-integer value in units of $\kappa_0= \pi^2 k_B^2 T /3 h$ ($T$ is temperature, $h$ the Planck constant, $k_B$ the Boltzmann constant). 
Such values were recently observed in a quantum-Hall system and a magnetic insulator. 
We show that nontopological ``thermal metal'' phases that form due to quenched disorder may disguise as non-Abelian phases by well approximating the trademark quantized thermal Hall response. 
Remarkably, the quantization here \textit{improves} with temperature, in contrast to fully gapped systems. 
We provide numerical evidence for this effect and discuss its possible implications for the aforementioned experiments.
\end{abstract}

\maketitle

{\bf Introduction}---Measurements of the electronic or thermal Hall effect are powerful experimental techniques for identifying topologically ordered phases and their fractional quasiparticles \cite{Kane1997}.
An electronic Hall conductance sharply quantized to a noninteger value (in units of $e^2/h$, with $e$ the electron charge) is intimately related to the existence of fractionally charged quasiparticles \cite{Arovas1984}. 
Similarly, a quantized noninteger thermal Hall conductance $\kappa_{xy}$ reflects excitations with non-Abelian braiding properties \cite{Moore1991,ReadRezayi1999, Read2000, Cappelli2002, Bernevig2008, Gromov2015}. 
While the electronic Hall effect has been routinely measured for several decades \cite{Klitzing1980}, precise measurements of the thermal Hall effect in solid-state systems have been achieved only recently~\cite{Jezouin2013, Banerjee2017, Banerjee2018, Kasahara2018, Srivastav2019}. 
Remarkably, experiments on two completely different systems found half-integer values, indicative of so-called Ising anyons. 
The first is a two-dimensional electron gas in a perpendicular magnetic field at filling factor $\nu=5/2$ with a thermal Hall conductance $\kappa_{xy}=5/2$ \cite{Banerjee2018}. 
The second is the  magnetic insulator $\alpha$-RuCl$_3$, where $\kappa_{xy}=1/2$ per layer was measured in an applied magnetic field \cite{Kasahara2018}. 

The level of quantization observed in the thermal Hall measurements is significantly below the one in their electronic analogs. 
This may be due to heat leakage from the measured system to its environment, e.g., via phonons.
Our work focuses on an additional, intrinsic property that may be particularly relevant to situations with half-integer $\kappa_{xy}$ under experimental conditions, which necessarily include sample imperfections, i.e., disorder \cite{Mross2018, WangHalperin2018, Biao2018}. 
In the context of both the quantum Hall effect and $\alpha$-RuCl$_3$, the half-integer $\kappa_{xy}$ results from topological $p_x\pm ip_y$ pairing \cite{Moore1991, Kitaev2006} (and consequently chiral edge Majoranas) of \textit{emergent} fermions: composite fermions and spinons, respectively. 
In both cases, the fermions are minimally coupled to an emergent gauge field, which acquires a Higgs mass and thereby eliminates the phase mode of the emergent-fermion superconductor. 
The effective low-energy theory thus falls into class D in the Altland-Zirnbauer classification (AZ) \cite{Altland1997}, which permits a delocalized \textit{thermal metal} phase \cite{Cho1997, Senthil1998, Senthil2000, Read2000, Bocquet2000, Chalker2001, Mildenberger2007, Evers2008, Kagalovsky2010, Laumann2012} (see Supplemental Material for more background \cite{SM}). 
This property is sharply distinct from, e.g., the symmetry class A of electrons in the quantum Hall effect, which always localizes unless the system is at a topological phase transition \cite{Kramer1993}. 
The thermal metal also crucially differs from the metallic state formed from electrons that weakly antilocalize due to spin-orbit coupling \cite{Hikami1980, Asada2002, Asada2004, Obuse2007}: It exhibits delocalized states only at energy $E=0$ where particle-hole symmetry holds; at any nonzero energy it crosses over into class A at long length scales and all states localize.

In this work we demonstrate remarkable consequences of these  localization characteristics: As temperature increases, the longitudinal thermal conductance \textit{decreases} and vanishes for thermodynamically large systems. 
Concomitantly, the thermal Hall conductance becomes better quantized.
This somewhat counterintuitive behavior arises because thermal conductance is determined not solely by the delocalized $E=0$ states, but by all states in an energy window $\sim T$. 
Since almost all of these states are localized, the thermal conductance vanishes.
This localization behavior is similar to the one exhibited by electrons near an integer quantum Hall plateau transition. 
There, an infinitely sharp transition at $T \neq 0$ is predicted on the single-particle level, but interactions render the width of the transition finite \cite{Wang2000}. 
In the present case, interactions allow thermal metal behavior to persist over a finite temperature window. 
We will comment on the role of interactions towards the end of our discussion. 

{\bf Model and symmetries}---For concreteness, we frame our discussion in the language of the quantum Hall plateau at $\nu=5/2$. 
At energies below the charge gap, we model the system by a quadratic Bogoliubov--de Gennes Hamiltonian. 
We require that it exhibits two topological phases that are related by time-reversal symmetry (TRS) and permits a thermal metal phase. 
(TRS of composite fermions corresponds to particle-hole symmetry of electrons within a single Landau level \cite{Son2015}, which relates the Pfaffian and anti-Pfaffian phases \cite{Lee_particle-hole_2007, Levin_particle-hole_2007}.)

\begin{figure}[tb]
 \centering
 \includegraphics[width=1\columnwidth]{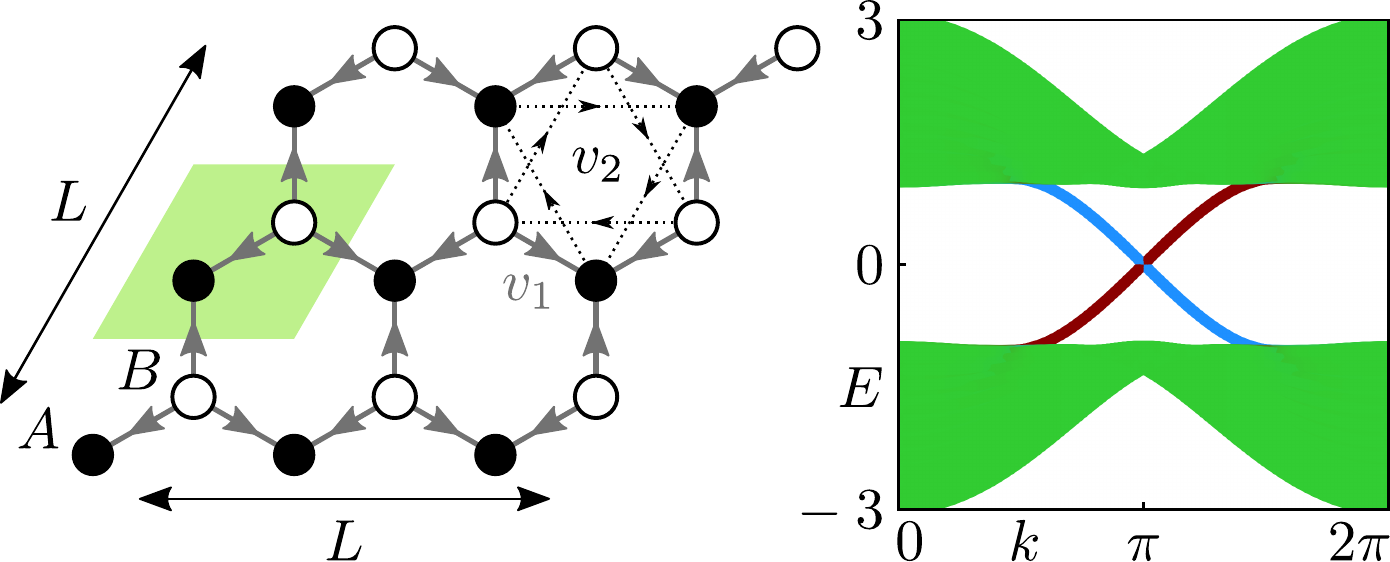}
 \caption{
 Left: The unit cell of the model (shaded area) contains two sites, indicated by solid and open circles, each hosting a single Majorana mode. 
 The signs of ($i$ times) the nearest-neighbor hopping $v_1$ and next-nearest-neighbor hoppings $v_2$ are indicated by the direction of arrows on the corresponding bonds. 
 Right: Band structure of the model in a ribbon geometry, infinite in the horizontal direction and consisting of 30 unit cells in the vertical direction, using $v_1=1$ and $v_2=-0.2$. 
 Bulk states are shown in green; states on the top and bottom edges are shown in red and blue, respectively.
 \label{fig:system}
 }
\end{figure}

A minimal model that satisfies these criteria is given by a honeycomb lattice with a single Majorana fermion per site as shown in Fig.~\ref{fig:system} \cite{Kitaev2006}. 
(Such a model also arises microscopically in the the Kitaev spin model below the vison gap. For a discussion of the thermal conductance at higher energies, see Ref.~\cite{Yang2020}) 
The Hamiltonian reads \begin{equation}\label{eq:kitaev_ham}
 H = i v_1 \sum_{\langle j,k \rangle} \gamma_j\gamma_k + i v_2 \sum_{\langle\langle j,k \rangle\rangle} \gamma_j\gamma_k,
\end{equation}
where $\gamma_j$ is a Majorana operator on site $j$, $ \langle j,k \rangle$ denotes directed bonds from B to neighboring A sites, and $\langle\langle j,k \rangle\rangle$ clockwise next-nearest neighbor bonds (see Fig.~\ref{fig:system}). 
We set $v_1=1$ throughout the following, expressing all energy scales relative to it. 

The Hamiltonian Eq.~\eqref{eq:kitaev_ham} obeys particle-hole (PH) symmetry. 
Writing $H=\boldsymbol{\gamma}^T {\cal H} \boldsymbol{\gamma}$, with $\boldsymbol{\gamma}$ a column vector of Majorana operators, the PH symmetry can be expressed as ${\cal H}=-{\cal H}^*$, such that the system belongs to class D. 
For $v_2=0$, the model features an additional TRS consisting of complex conjugation followed by a sign change of the wavefunction on one of the two sublattices, and thus falls into the symmetry class labeled BDI in the AZ classification. 
Here, the spectrum is gapless, with two linearly dispersing Majorana cones at momenta $K$ and $K'$.
A nonzero mass term $v_2$ results in a gapped phase with Chern number ${\cal C}=\text{sgn}(v_2)$. 

Notice that TRS, which is present for $v_2=0$, implies a vanishing thermal Hall conductance, while the corresponding particle-hole symmetry in the first excited Landau level requires the value $5/2$ \cite{footnote1}. 
Consequently, a ``background'' contribution of $5/2$ must be added to interpret results for the model system in the quantum Hall context, i.e., $\kappa^\text{QH}_{xy}=\kappa_{xy}+5/2$. 
The TRS breaking parameter $v_2$ describes either Landau level mixing or the deviations of $\nu$ from $5/2$.

{\bf Zero-temperature phase diagram}---We now introduce random hopping disorder and examine the localization properties near the topological phase transition of the clean system as a function of disorder strength $V$ and energy $E$. 
Specifically, we replace $v_i \to v_i + \delta v_i$ with $\delta v_1$ from a uniform distribution $[-V,V]$ and $\delta v_2$ independently from $[-V/10,V/10]$. 
We numerically compute the energy-dependent transmission probability $P(E)$ for a cylindrical $L\times L$ system with zigzag edges. 
(See Supplemental Material for details \cite{SM}). 

\begin{figure}[tb]
 \includegraphics[width=\columnwidth]{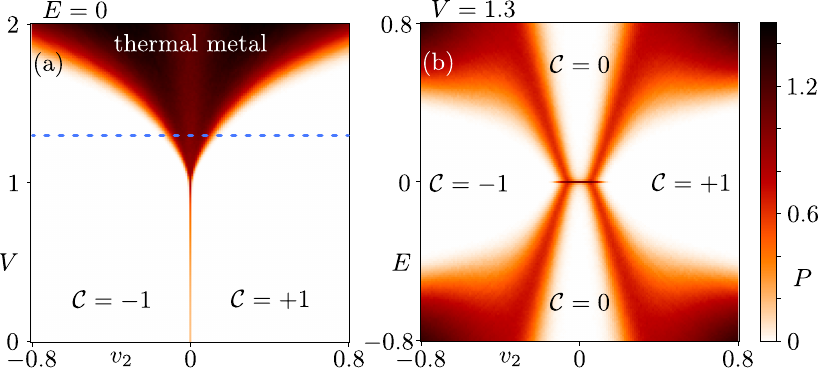}
 \caption{The average transmission of a $80\times80$ unit cells system, computed using 1000 disorder realizations. 
(a) The horizontal axis is $v_2$ and the vertical axis is disorder strength~$V$. 
 The topological transition between phases with Chern numbers ${\cal C}=\pm 1$ evolves into a thermal metal phase with increasing disorder strength. 
 The dashed blue line indicates $V=1.3$. 
(b) For fixed $V=1.3$, the average transmission is plotted as a function of $v_2$ and energy~$E$. 
 Away from the particle-hole symmetric line, $E=0$, the thermal metal phase disappears; it is replaced by topologically trivial insulators (${\cal C}=0$).
 \label{fig:pd_2term}
 }
\end{figure}

In Fig.~\ref{fig:pd_2term}(a) we show the transmission at zero energy as a function of $v_2$ and disorder strength. 
At weak disorder, the insulating ${\cal C}=\pm 1$ phases remain separated by a direct plateau transition, see also discussion in Refs.~\cite{Cho1997, Chalker2001, Mildenberger2007, Evers2008, Kagalovsky2010}. 
When $V\gtrsim 1$, however, a delocalized phase develops around $v_2=0$. 
This is a disorder-induced thermal metal where $P(E=0)$ increases logarithmically with system size (see Supplemental Material \cite{SM}). 
Further increasing the disorder strength enlarges the thermal metal region.
In Fig.~\ref{fig:pd_2term}(b), we plot the energy-dependent transmission probability at fixed disorder strength $V=1.3$, where the thermal metal is well developed. 
We observe that the metal is only present at $E=0$. 
This is the expected result based on weak antilocalization in the PH symmetric case, $E=0$, and weak localization for any $E\neq 0$. 
The latter case features direct insulator-to-insulator transitions, with a trivial, $C=0$ phase between the topological ${\cal C}=\pm1$ phases.

We have further observed that the crossover regions in Fig.~\ref{fig:pd_2term}(b) (the regions with nonzero transmission) shrink with increasing system size and grow with $\delta v_2$ disorder. 
The former is required due to the absence of metallic phases in class A, i.e., for $E \neq 0$. 
To understand the latter, recall that for $v_2=\delta v_2=0$ the model is in class BDI, which only permits a critical-metal phase (without antilocalization, see Supplemental Material \cite{SM}), distinct from the class-D thermal metal \cite{Evers2008}. 
Gradually introducing $\delta v_2$ disorder allows the formation of a growing thermal metal phase at $E=0$ and, in finite systems, a corresponding crossover at $E\neq 0$. 
For the numerically accessible system sizes, we find that the $C=0$ insulator fully disappears into a smooth crossover for approximately equal $v_1$ and $v_2$ disorder. 

At $E\neq 0$ the system is in class $A$ and thus always localizes. 
For small energies, the localization length is determined by the crossover between weak antilocalization of class D and weak localization of class A \cite{Mildenberger2007}. 
At distances below the diffusion length $L_E=\sqrt{D/E}$, with $D$ the diffusion constant, the interference of electron and hole trajectories gives rise to a logarithmic increase of the conductance. 
At scales above $L_E$, this interference is suppressed; the system behaves as a class-A conductor, which tends to localize. The localization length $\xi(E)$ can be computed as for the case of spin-orbit coupled electrons in a weak magnetic field, which features a similar crossover between weakly antilocalizing class AII and weakly localizing class A \cite{Lerner1995}. 
We find
\begin{align}
 \xi(E)= \frac{\ell_0}{\sqrt{E}} \exp \left( \frac{1}{4} \ln^2 \frac{1}{E} \right), \label{eqn.sigmascaling}
\end{align}
with $\ell_0$ the mean free path (see Supplemental Material \cite{SM}). 
Both $L_E,\xi(E)$ depend on energy and diverge with decreasing energy, with asymptotically $\xi(E)\gg L_E$. 
In a finite-size system it is further useful to define two crossover energy scales: $E_L\propto L^{-2}$, for which $L_E=L$, and $E_c$, for which $\xi=L$.
States with energies below $E_L$ are weakly antilocalizing, states with energies between $E_L$ and $E_c$ are characterized by weak localization, whereas strong localization sets in above $E_c$.

\begin{figure}[tb]
 \includegraphics[width=1\columnwidth]{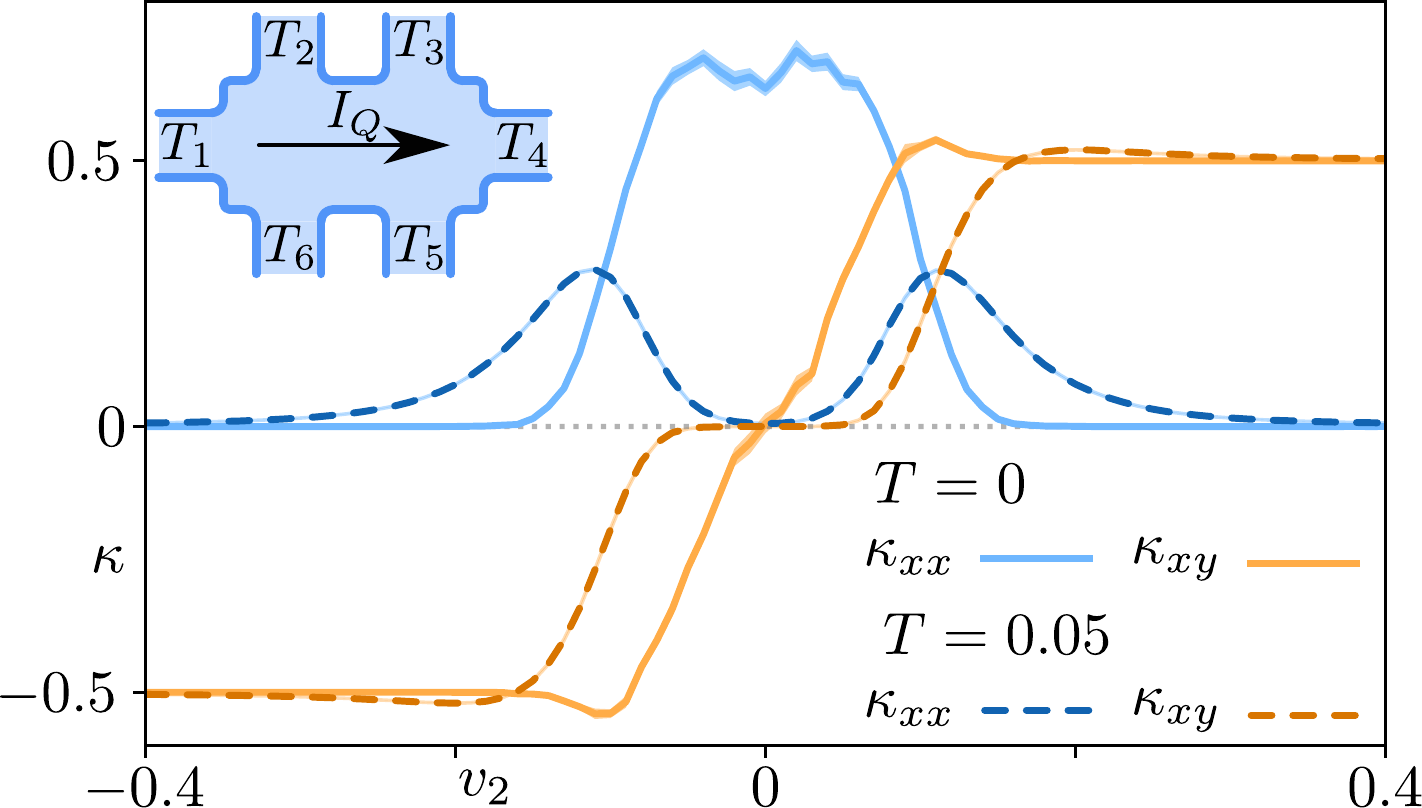}
 \caption{The dimensionless longitudinal and transverse thermal conductances (blue, orange) are computed in a six-terminal geometry (inset) at $T=0$ and $T=0.05$ (solid, dashed). 
 We use a rectangular system composed of 80 zigzag chains in the vertical direction and 160 hexagonal plaquettes in the horizontal direction, and average over 1000 disorder realizations ($V=1.3$), with line thickness indicating the error bars. 
 The thermal metal phase present in the $v_2=0$ region of the plot is converted into an insulating phase at nonzero temperature, leading to a quantized plateau in both longitudinal and transverse conductance. 
 Notice the smaller error bars in the finite-temperature curves, which occur because the energy integral Eq.~\eqref{eq:condmatrix} provides additional disorder averaging. \label{fig:6term}}
\end{figure}

{\bf Finite temperature effects}---The above discussion and the results presented in Fig.~\ref{fig:pd_2term} suggest that the dimensionless thermal conductance tensor $\kappa$ changes qualitatively as temperature is swept past either crossover scale. 
(In infinite systems $E_c=E_L=0$.) 
To test this expectation, we compute $\kappa_{xx}$ and $\kappa_{xy}$ in a six-terminal transport geometry, where both can be obtained in the same numerical simulation. 
Terminals are numbered from 1 to 6, as shown in the inset of Fig.~\ref{fig:6term}. 
The scattering matrix has a $6\times 6$ block structure, with blocks $\mathfrak{s}_{ij}$ containing the probability amplitudes for transmission from lead $i$ to lead $j$.

All reservoirs are kept at equal chemical potentials, and terminals 1 and 4 are temperature-biased as $T_{1,4}=\pm \Delta T/2$ relative to the base temperature $T$. 
Consequently,  a heat current $I_Q=I_1=-I_4$ flows through the system. 
The remaining terminals are connected to thermometers, such that $I_{2,3,5,6}=0$. 
Setting $h=k_B=1$, temperatures and heat current are related via \cite{Long2011}
\begin{equation}\label{eq:IvsT}
 \begin{pmatrix}
 I_Q \\
 0 \\
 0 \\
 -I_Q \\
 0 \\
 0 
 \end{pmatrix} = M
 \begin{pmatrix}
 \Delta T/2 \\
 T_2 \\
 T_3 \\
 - \Delta T /2 \\
 T_5 \\
 T_6
 \end{pmatrix},
\end{equation}
where the matrix $M$ is given by
\begin{equation}\label{eq:condmatrix}
 M_{ij} = \int_0^\infty \frac{E^2}{T} \left( -\frac{\partial f(E, T)}{\partial E} \right)\left[\delta_{ij}N_j- {\rm tr}(\mathfrak{s}_{ij}^\dag \mathfrak{s}^\pd_{ij}) \right] dE,
\end{equation}
with $\delta_{ij}$ denoting the Kronecker delta, $N_j$ the number of modes in lead $j$, and $f(E,T)=1/(1+e^{E/T})$ the Fermi-Dirac distribution. 
Notice that the scattering matrix depends on energy, which results in a nontrivial temperature dependence of $M$. 
We numerically compute the matrix $M$ for each disorder realization, and insert it into Eq.~\eqref{eq:IvsT} to calculate the elements of the thermal resistance tensor, $R_{xx}$ and $R_{xy}$. 
We define $R_{xx} =(T_2-T_3)/ I_Q$, whereas for $R_{xy}$ we average over the temperature drop between terminals 2 and 6 and between terminals 3 and 5: $R_{xy} = (T_2-T_6 + T_3-T_5)/ (2 I_Q)$ to reduce geometric effects. 
The dimensionless conductance tensor $\kappa$ is then obtained by inverting $R$ and dividing by $\kappa_0$. 
In Fig.~\ref{fig:6term} we plot the disorder-averaged components $\kappa_{xx}$ and $\kappa_{xy}$ at $T=0$ (solid lines) and $T=0.05$ (dashed lines). We observe that at zero temperature the transition between ${\cal C}=\pm 1$ phases in which $\kappa_{xy}=\pm 0.5$ occurs via an intermediate thermal metal phase. This is signaled by a large peak in $\kappa_{xx}$ and a nonquantized $\kappa_{xy}$. At finite temperature the thermal metal peak is replaced by an intermediate plateau of $\kappa_{xx/xy} \approx 0$ in the small $|v_2|$ region of the plot. At large $|v_2|$, however, the dimensionless thermal conductance remains qualitatively unchanged, with $\kappa_{xy}\to \pm0.5$ and $\kappa_{xx}\to 0$, since the insulating phases with ${\cal C}=\pm 1$ extend to an energy range larger than temperature (see Fig.~\ref{fig:pd_2term}).

\begin{figure}[tb]
 \includegraphics[width=\columnwidth]{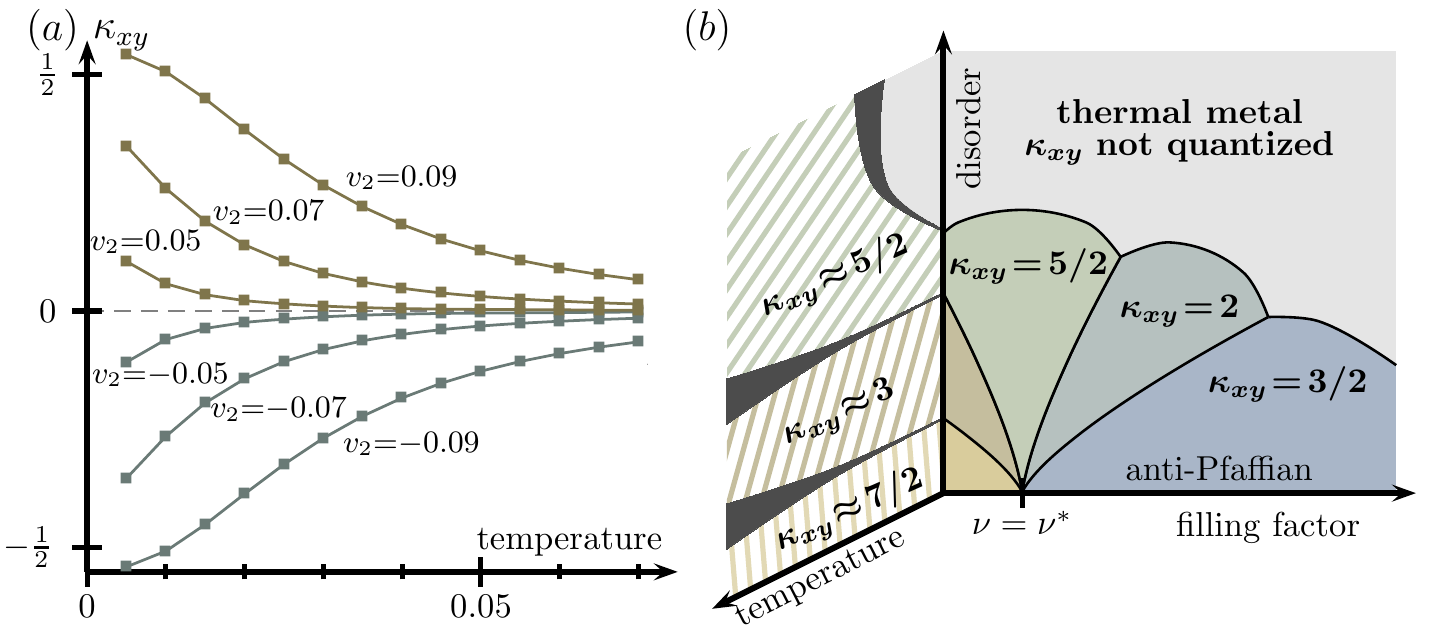}
 \caption{(a) The numerically computed thermal conductance shows order-unity variations with $v_2$ at low temperatures. 
 At higher temperatures it approaches the time-reversal symmetric value $\kappa_{xy}=0$ with very weak dependence on $v_2 \in[-0.07,0.07]$. 
 (b) Proposed finite-temperature phase diagram of $\nu\approx 5/2$ quantum Hall states. 
 The thermal metal extends to finite temperatures due to residual interactions between neutral quasiparticles. 
 At higher temperatures, but still below the charge gap, the mechanism discussed here sets in. 
 Near $\nu^*$ it results in an large region where $\kappa_{xy}$ approaches the particle-hole symmetric value $5/2$. 
 Hatched regions denote quantization of $\kappa_{xy}$ that is better than a threshold value, say 1$\%$. 
 The approximate plateaus are separated by relatively sharp crossovers as shown in Fig.~\ref{fig:6term}. 
 \label{fig:conc}
 }
 \label{finitetempphases}
\end{figure}

To understand the observed behavior, notice that temperature-dependent factor in the integrand of Eq.~\eqref{eq:condmatrix} is sharply peaked at energies around $T$ and becomes a delta function at zero temperature. 
For temperatures below the crossover energy scale $E_c$, the main contribution to the integral comes from extended states and $\kappa_{xx}$ is large while $\kappa_{xy}$ is approximately linear in $v_2$. 
The low-temperature regime subdivides into $T<E_L$, where $\kappa_{xx}$ is determined by the metallic transmission probability $P(E \approx 0)$ and $E_L < T < E_c$, where weak localization modifies this result (see Supplemental Material \cite{SM}). 
At temperatures above $E_c$, strong localization becomes operative and leads to $\kappa_{xx}\rightarrow 0$, as we observe numerically. 
This insulating behavior is accompanied by an emergent intermediate plateau in $\kappa_{xy}$, as shown in Figs.~\ref{fig:6term} and  \ref{fig:conc}(a). 

{\bf Effect of interactions}---Both the numerical simulations and the $\sigma$-model treatment were performed within quadratic fermionic Hamiltonians, where transport is fully phase coherent. 
We now qualitatively discuss how interactions change these results, focusing on the fractional quantum Hall effect at $\nu=5/2$. 
We consider a thermodynamically large system, $L\rightarrow \infty$, and focus on phase-breaking of the emergent fermions. 
The presence of phonons leads to another channel for heat transport, which is expected to follow a power law of $\kappa^{ph}\propto T^5$ \cite{Ziman2001}. 
Its contribution is suppressed at low temperatures, as indeed observed in the experiment \cite{Banerjee2018, Kasahara2018}.

The phase-breaking interactions of an interfering particle with its environment introduce a dephasing length $L_\phi$ \cite{Imry1997}, which diverges with decreasing temperature, presumably following a power law. 
This length should be compared to the diffusion length $L_E$ and the localization length $\xi(E)$ [defined near Eq.~\eqref{eqn.sigmascaling}]. 
The relevant energy is the temperature, since the chemical potential is zero. 

For a clean sample at high temperature, the dephasing length is presumably the shortest scale, and the system's conductance takes the antilocalizing class-D form, where $\kappa_{xx}\propto \log{L_\phi/\ell_0}$. 
As the temperature is lowered, we will have $L_\phi>L_E$ and patches of size $L_\phi$ become large enough for each to manifest a tendency to localization as a system of class A. 
Thus, in this regime, the conductance decreases with decreasing temperature, with its maximum $\log{(D/T\ell_0^2 )}$ being attained when $L_\phi\approx \sqrt{D/T}$. 
If phase breaking is sufficiently weak, there will be an intermediate range of temperatures in which $L_\phi>\xi$ and we expect a conductance much smaller than unity. 
The charge neutrality of the Majorana particles suggests that such a regime may indeed exist. 
In any case, in the zero-temperature limit the energy dependence of $\xi$ [see Eq.~\eqref{eqn.sigmascaling}] makes it larger than~$L_\phi$ and the system becomes metallic. 

{\bf Conclusion}---We have shown that in systems whose effective low-energy theory falls into symmetry class D, increasing temperature can \textit{enhance} quantization of the thermal Hall response. 
Moreover, new plateaus of quantized $\kappa_{xy}$ that are absent in the low-temperature limit may emerge and exhibit near-perfect quantization. 
These findings have direct implications for the interpretation of the measured half-integer values of $\kappa_{xy}$ in Refs.~\cite{Banerjee2018, Kasahara2018}. 

In the context of the $\nu=5/2$ plateau, it is not settled whether the observed thermal Hall conductance represents a bulk property or is due to incomplete equilibration between edge states \cite{Simon2018, Feldman2018cSimon, Simon2018rFeldman, Ma2019, Simon2020, Asasi2020, Park2020}. 
In the former case, the PH-Pfaffian with topologically quantized $\kappa_{xy}=5/2$, a disorder-induced topological phase with the same quantized value, or a thermal metal where $\kappa_{xy}$ is not strictly quantized are all possible~\cite{Mross2018, WangHalperin2018, Biao2018}. 
If particle-hole symmetry were present, it would constrain the Hall response of the thermal metal to be the same as that of PH-Pfaffian. 
In reality, this symmetry is broken due to relatively strong mixing between Landau levels -- the Coulomb energy is comparable to the Landau level splitting in the relevant experiments. 
Our work shows how an approximately quantized thermal Hall response can emerge with increasing temperature out of a zero-temperature thermal metal whose response is nonquantized. 
We thus propose that the zero-temperature phase diagram of the quantum Hall state near $\nu=5/2$ introduced in Refs.~\cite{Mross2018, WangHalperin2018, Biao2018} extends to nonzero temperatures as shown in Fig.~\ref{finitetempphases}(b). 

More generally, our work emphasizes the importance of systematically measuring the temperature dependence of the approximately quantized $\kappa_{xy}$. 
(In the case of $\alpha-$RuCl$_3$ the thermal Hall measurements were performed at moderate temperatures of around 4 K.) 
Theoretically, a careful analysis of dephasing could help determine whether an underlying thermal metal is possible or if the measured $\kappa_{xy}$ reflects a topological phase. 

\begin{acknowledgments}
It is a pleasure to thank Olexei Motrunich, Anton Akhmerov, and Moty Heiblum for illuminating discussions on this topic, and Ulrike Nitzsche for technical assistance. 
This work was partially supported by the Deutsche Forschungsgemeinschaft (DFG, German Research Foundation) under Germany's Excellence Strategy through the W\"{u}rzburg-Dresden Cluster of Excellence on Complexity and Topology in Quantum Matter -- \emph{ct.qmat} (EXC 2147, project-id 390858490), through CRC/Transregio 183, and through grants EI 519/7-1 and MI 658/10-1, by the ERC under the European Union’s Horizon 2020 research and innovation programme (grant agreement LEGOTOP No 788715), the BSF and NSF (2018643), the ISF (1866/17), the ISF Quantum Science and Technology (2074/19), and by the German-Israeli Foundation (grant I-1505-303.10/2019).

\end{acknowledgments}

\bibliography{References}

\begin{thebibliography}{52}
\bibitem [{foo()}]{footnote1}%
  \BibitemOpen
  \href@noop {} {\bibinfo  {journal} {Neither symmetry is related to the
  intrinsic {PH} symmetry of {BdG} {H}amiltonians, which does not impose any
  such constraint}\ }\BibitemShut {NoStop}%
\end{thebibliography}

\end{document}


\title{Supplemental Material to: ``Temperature enhancement of thermal Hall conductance quantization''}

\author{I. C. Fulga}
\affiliation{IFW Dresden and W{\"u}rzburg-Dresden Cluster of Excellence ct.qmat, Helmholtzstrasse 20, 01069 Dresden, Germany}

\author{Yuval Oreg}
\affiliation{Department of Condensed Matter Physics, Weizmann Institute of Science, Rehovot 76100, Israel}

\author{Alexander D. Mirlin}
\affiliation{Institute for Quantum Materials and Technologies, Karlsruhe Institute of Technology, 76021 Karlsruhe, Germany}
\affiliation{Institut f\"{u}r Theorie der Kondensierten Materie, Karlsruhe Institute of Technology, 76128 Karlsruhe, Germany}
\affiliation{L.\,D.~Landau Institute for Theoretical Physics RAS, 119334 Moscow, Russia}

\author{Ady Stern}
\affiliation{Department of Condensed Matter Physics, Weizmann Institute of Science, Rehovot 76100, Israel}

\author{David F. Mross}
\affiliation{Department of Condensed Matter Physics, Weizmann Institute of Science, Rehovot 76100, Israel}

\date{\today}
\begin{abstract}
In this Supplemental Material we provide some background on the thermal metal phase of class D systems, describe the transport geometry used to determine the thermal conductance, and show numerically obtained conductance scaling plots. 
In addition, we estimate energy dependence of the localization length both numerically and by using a nonlinear $\sigma$ model approach.
\end{abstract}
\maketitle

\section{Thermal metal}
\label{sec:thmetal}

In the Altland-Zirnbauer classification \cite{Altland1997}, class D refers to systems that break quasiparticle number conservation, time-reversal symmetry, as well as spin-rotation symmetry. Well-known representatives of this class are spin-triplet odd-parity superconductors, such as those showing chiral $p$-wave or $f$-wave pairing. Since charge and spin are not conserved, electrical and spin conductance are not suitable for distinguishing different phases in this symmetry class. Instead, their localization behavior is reflected in \emph{thermal transport}. Specifically, the dimensionless thermal conductance $\kappa_{xx}/\kappa_0$ vanishes in the $T\rightarrow 0$ limit for thermal insulators, but is nonzero in thermal metals.

Thermal metals may form in class D due to quantum interference effects that cause weak antilocalization of quasiparticles. By contrast, symmetry classes that feature weak localization (such as class A) cannot exhibit such a phase. The effect of interference on the conductance can be conveniently encoded in the scaling of the inverse conductance $t=1/(\pi g)$ with systems size $L$ as $
- \frac{dt}{d \ln L} = \beta(t)$. For the symmetry classes discussed in this work, the $\beta$ functions in two dimensions read \cite{Evers2008}
\begin{align}
\beta_\text{D}(t) &= t^2~, \label{eq:betaD}\\
\beta_\text{AII}(t) &= t^2/2~,\label{betaaii}\\
\beta_\text{A}(t) &=- t^3/2~,\label{eq:betaA}\\
\beta_\text{BDI}(t) &= 0~\label{betabdi}.
\end{align}
Eqs.~\eqref{eq:betaD}-\eqref{eq:betaA} hold to the leading nontrivial order of perturbation theory in $t$, whereas Eq.~\eqref{betabdi} holds to all orders of the perturbation theory.

For the classes D and AII, the $\beta$ function is positive. The conductance increases logarithmically with system size, i.e., exhibits (thermal) metallic behavior. For class A---no symmetries other than charge conservation---it is negative. Here, the conductance away from the quantum Hall transition decreases, leading to insulating behavior. Finally, when a system in class D is endowed with an extra chiral symmetry, its symmetry class changes from D to BDI, and quantum interference effects cancel to yield a critical (system-size independent) conductance.

\section{Transport geometry}
\label{sec:geometry}

All numerical simulations are performed using the Kwant code \cite{Groth2014}. 
In the two-terminal geometry, we use a square shaped system consisting of $L\times L$ unit cells, with periodic boundary conditions in the horizontal direction. 
Ideal leads are attached to the top- and bottom-most unit cells of the system and are modeled as ``doped'' Kitaev models. 
The lead Hamiltonian has the same form as in the main text (with $v_2=0$), except that it is considered to describe spinless, complex fermions instead of Majorana modes. 
The doping then corresponds to a global shift of the Fermi level: ${\cal H} \to {\cal H} + 0.5v_1$, which breaks particle-hole symmetry. 
These leads are connected to fermionic reservoirs that are held at identical electro-chemical potential but different temperatures. 
The resulting scattering matrix,
\begin{equation}\label{eq:smatrix}
S = \begin{pmatrix}
\mathfrak{r} & \mathfrak{t}' \\
\mathfrak{t} & \mathfrak{r}' 
\end{pmatrix},
\end{equation}
is composed of blocks $\mathfrak{r}^{(\prime)}$ and $\mathfrak{t}^{(\prime)}$ containing the probability amplitudes for lead modes to be back-reflected or transmitted across the system, respectively. 
As mentioned in the main text, this allows us to determine the transmission probability through the system as $P(E)={\rm tr} (\mathfrak{t}^\dag \mathfrak{t})$. 
Notice that this total transmission probability includes a sum over transmissions of a fermion emanating from  one reservoir to a fermion that is absorbed in another reservoir. 
The fermions can be either electrons or holes, as all processes transfer energy between the reservoirs. 
The use of leads and reservoirs here should be viewed as a computational vehicle allowing us to probe theoretically the localization properties of the system's bulk. 
The modeling of the hot and cold reservoirs and the leads in actual experiments are more involved, but that should not affect significantly the bulk properties and their temperature dependence that we study here.

To determine the longitudinal thermal conductance in the two-terminal geometry, we assume that the fermionic reservoirs are kept at the same chemical potential. 
Instead, the reservoirs are temperature biased by an amount $\pm\Delta T/2$ relative to the temperature $T$ of the superconductor. 
As such, a heat current $I_Q$ flows through the device, leading to a longitudinal thermal conductance $\kappa_0\kappa_{xx}=I_Q/\Delta T$, which in linear response reads
\begin{equation}\label{eq:cond}
 \kappa_{xx}(T) = \frac{1}{T\kappa_0}\int_0^\infty E^2 \left( -\frac{\partial f(E, T)}{\partial E}  \right)P(E)dE.
\end{equation}
Here, $f(E,T)=\left( 1+\exp(E/T) \right) ^{-1}$ is the Fermi function at temperature $T$, relative to the Fermi level $E_F=0$, and we have set $h=k_B=1$. 
In the limit of zero temperature, $T\to 0$, Eq.~\eqref{eq:cond} yields a dimensionless conductance $\kappa_{xx} = P(E=0)$.

In the six terminal geometry, we consider a rectangular system composed of 80 zig-zag chains in the vertical direction and 160 hexagonal plaquettes in the horizontal direction. 
The six leads are positioned symmetrically with respect to each edge of the system, as sketched in the inset of Fig.~3 of the main text. 
In the six terminal setup, the lead Hamiltonian is formed out of decoupled Kitaev chains at their critical point (hopping strength equal to $v_1$), with each chain connecting to one site of the system.

\section{Conductance scaling}
\label{sec:cond}

\begin{figure}[tb]
 \includegraphics[width=\columnwidth]{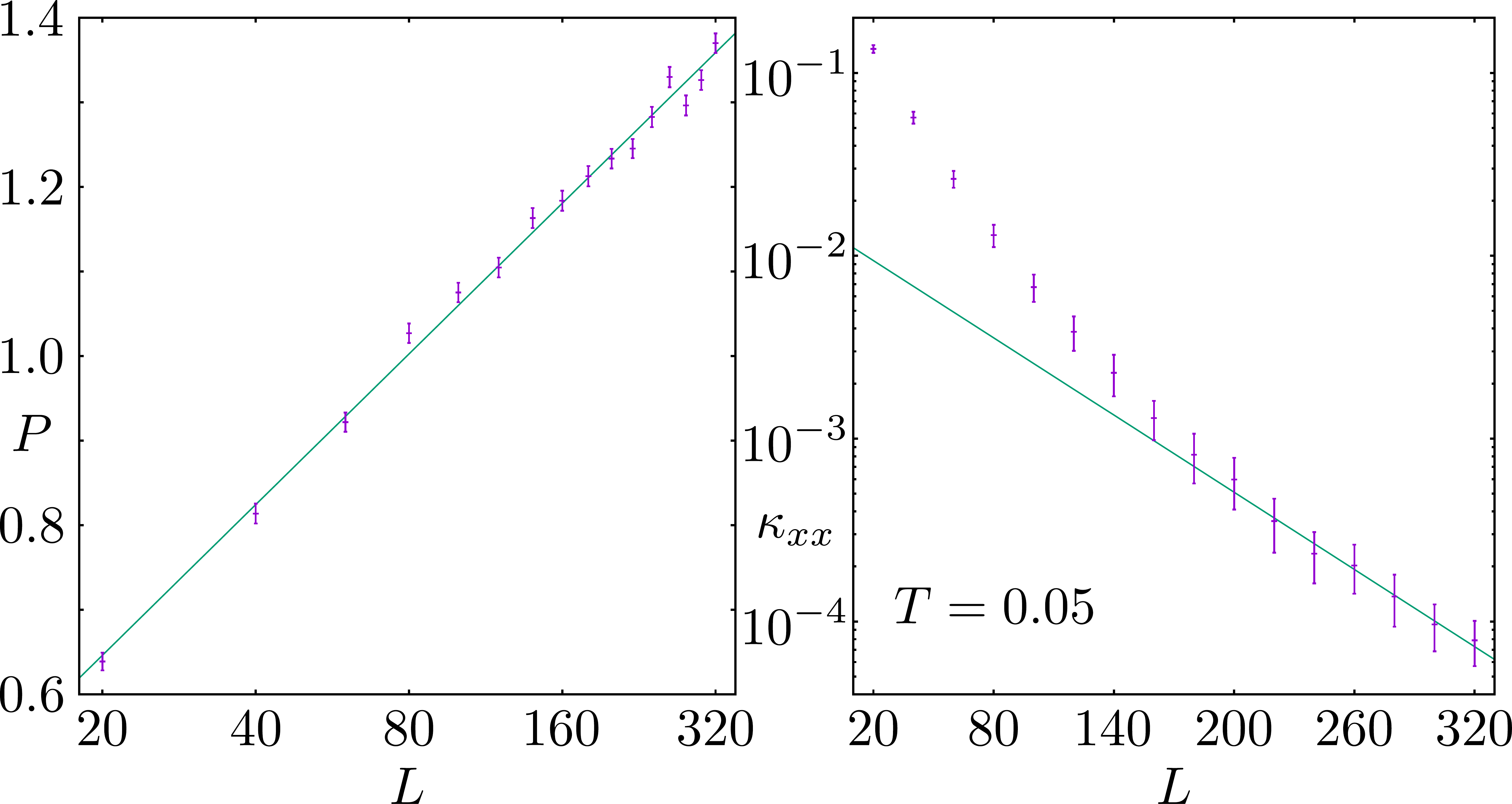}
 \caption{Left panel: the average transmission at $E=0$ is plotted versus system size setting $v_2=0$ and $V=1.3$. 
 Notice the log scale on the horizontal axis. 
 The solid line shows a logarithmic increase, expected for a thermal metal. 
 Right panel: the average dimensionless conductance is plotted versus system size for $T=0.05$, $v_2=0$, and $V=1.3$. 
 Notice the log scale on the vertical axis. 
 The solid line shows an exponential decay. 
 In both panels, each point was obtained by averaging over 1000 disorder realizations.
 \label{fig:cond_vs_L_vary_T}
 }
\end{figure}

As shown in the left panel of Fig.~\ref{fig:cond_vs_L_vary_T}, in the limit of zero temperature the thermal metal phase is characterized by a logarithmic scaling of conductance with system size, as expected for weak anti-localization. 
The fit to $\ln L$ yields a slope $\sim0.26$, close to the predicted value of $1/\pi$ \cite{Evers2008}. 
At finite temperatures, however, we observe that the conductance decreases with system size instead, as shown in the right panel of Fig.~\ref{fig:cond_vs_L_vary_T}. 
Since the disorder average commutes with the energy integral of Eq.~\eqref{eq:cond}, we determine the finite temperature $\kappa_{xx}$ by first averaging the transmission probability over disorder, and then numerically evaluating the energy integral. 
At nonzero $T$, the error bars of the conductance are obtained by re-computing the integral using transmission probabilities which are all one error bar above or one error bar below the mean.

\begin{figure}[tb]
 \includegraphics[width=1\columnwidth]{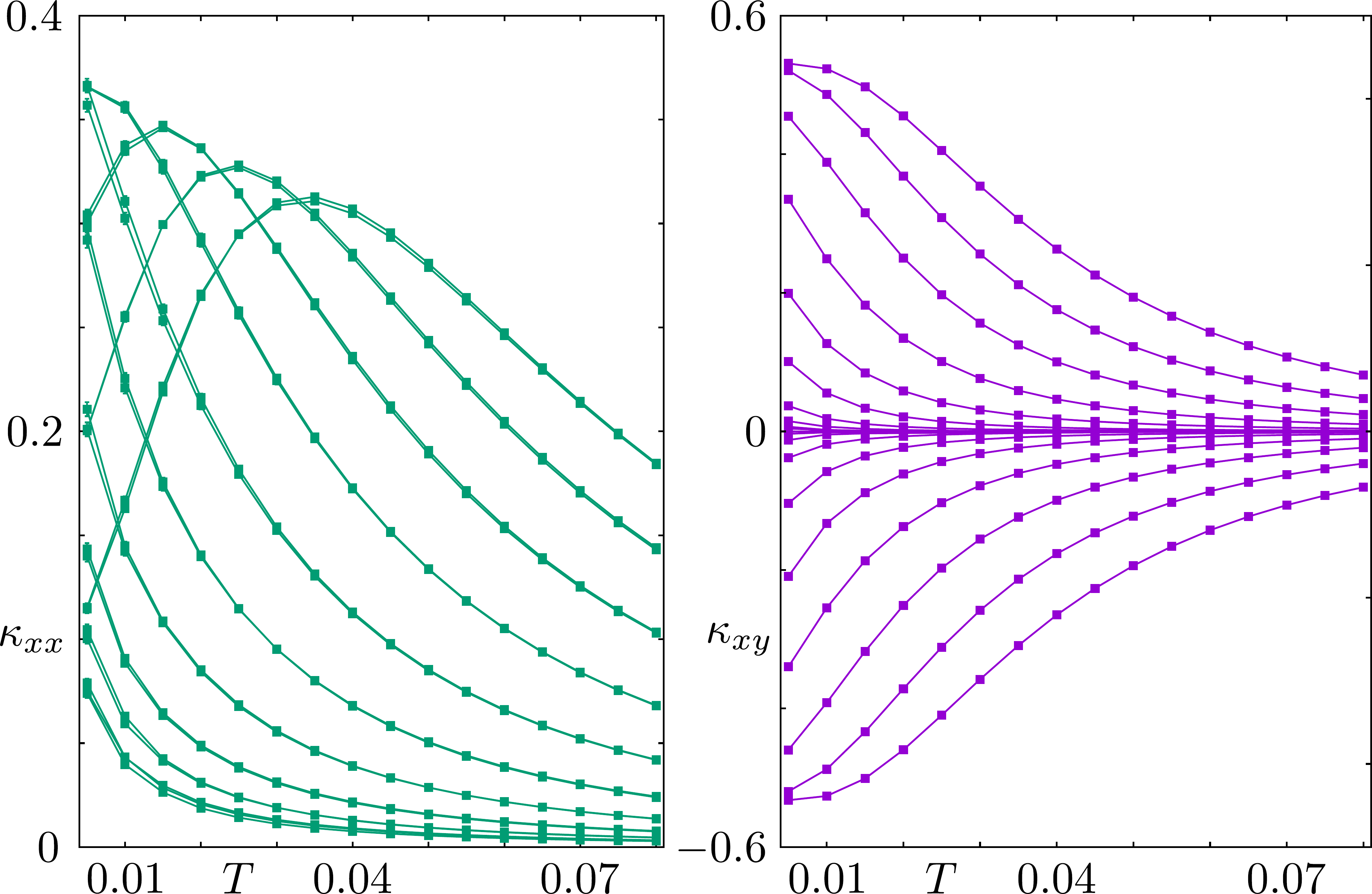}
 \caption{Average longitudinal (left) and transverse (right) dimensionless conductance plotted as a function of temperature. 
 We use the same system size as in Fig.~3 of the main text, setting $v_1=1$ and $V=1.3$. 
 Each point is obtained by averaging over $1000$ disorder realizations. 
 Each curve corresponds to a different value of $v_2$, from $-0.1$ to $0.1$ in steps of $0.01$, and the solid lines are guides to the eye. 
 In the left panel, larger values of $\kappa_{xx}$ correspond to larger absolute values $|v_2|$, whereas in the right panel the larger $\kappa_{xy}$ corresponds to larger $v_2$. 
 \label{fig:cond_vs_T_fix_L}
 }
\end{figure}

Keeping a fixed system size, we observe that for small $v_2$, both the longitudinal and the transverse dimensionless conductance decrease as a function of temperature, consistent with the formation of a finite-temperature plateau. 
This behavior is shown in Fig.~\ref{fig:cond_vs_T_fix_L}, using the six terminal geometry discussed earlier. 
As mentioned in the main text, conductance quantization improves with increasing temperature.

\section{Localization length}
\label{sec:loc_len}

In the following, we estimate the energy dependence of the localization length in the framework of a nonlinear $\sigma$ model \cite{Evers2008}. 
At low energies, for which the system size is much smaller than the localization length but larger than the mean-free path, $\ell_0$, the renormalization group (RG) flow of the $\sigma$ model proceeds as in class D, with the beta function in Eq.~\eqref{eq:betaD}. The coupling constant $t=1/(\pi P)$ is proportional to the inverse of the dimensionless conductance in the zero temperature limit, which is equivalent to the transmission probability $P$, computed in the cylinder geometry. 
This variable is usually denoted by $g$ in $\sigma$ model calculations, but here we refer to it as $P$ to be consistent with previous notation. 
The solution of Eq.~\eqref{eq:betaD} is a logarithmically growing transmission probability, 
\begin{equation}\label{eq:condD}
P(L) = {\rm const} + \frac{1}{\pi} \ln\frac{L}{\ell_0},
\end{equation}
consistent with the positive scaling obtained numerically in Fig.~\ref{fig:cond_vs_L_vary_T}, left panel.
To relate the above transmission probability to an energy scale, we use the RG equation for the second coupling constant, $\varepsilon$, whose bare value is the energy, $E$. 
This flow equation reads
\begin{equation}\label{eq:flowE}
 \frac{d\varepsilon}{d\ln (L/\ell_0)} = (2+t)\varepsilon.
\end{equation}
Using the fact that the thermal metal transmission probability is large, meaning $t\ll 1$, we solve Eq.~\eqref{eq:flowE} to obtain $\varepsilon\simeq E (L/\ell_0)^2$. 
The class D RG flow towards larger $L$ is therefore stopped by the energy, when the coupling constant $\varepsilon\sim 1$. 
After this point, the flow continues as in class A, corresponding to weak localization. 
Using that $\varepsilon\simeq E (L/\ell_0)^2$ is of order unity at this crossover point, we can estimate the crossover length scale at which the class A RG starts as 
\begin{equation}\label{eq:Lc}
 L_c = \ell_0 \left( \frac{1}{E} \right)^{1/2},
\end{equation}
corresponding to a transmission probability
\begin{equation}\label{eq:gc}
 P_c\equiv P(L_c) = {\rm const} + \frac{1}{2\pi} \ln \left(\frac{1}{E} \right).
\end{equation}
Starting from this value, the flow equation is now given by the negative beta function of class A [see Eq.~\eqref{eq:betaA}, signaling localization. 
Solving Eq.~\eqref{eq:betaA} gives
\begin{equation}\label{eq:condA}
 P(L) = \sqrt{P_c^2 - \frac{1}{\pi^2}\ln\frac{L}{L_c}}.
\end{equation}
The transmission probability now decreases as a function of system size from its initial value, $P_c$. 
To estimate the localization length, we use the approach of Ref.~\cite{Lerner1995}, which defines $\xi$ as that length for which the transmission probability flows to a constant of order unity, $P(\xi)\sim 1$. 
If we neglect the constant term in Eq.~\eqref{eq:gc}, meaning that we consider the logarithm to be large, we obtain
\begin{align}\label{eq:xiofE}
 \xi(E) \simeq & L_c \exp\left(\pi^2P_c^2\right) \nonumber\\
 = & \frac{\ell_0}{\sqrt{E}} \exp \left( \frac{1}{4} \ln^2 \frac{1}{E} \right).
\end{align}
The localization length has a sub-leading behavior which is power-law, $E^{-1/2}$. 
However, its asymptotic form is between a power law and an exponential: It grows like exponent of logarithm square. 
This result is consistent with a finite temperature conductance that decreases with system size, as shown in Fig.~\ref{fig:cond_vs_L_vary_T}, right panel.

The localization length formula shown in the first line of Eq.~\eqref{eq:xiofE} is consistent with discussion in Ref.~\cite{Lerner1995} on how the localization length evolves when a magnetic field is added to an initially time-reversal symmetric, spin-orbit coupled system (class AII). 
There too, there was a crossover to class A in the RG flow, leading to a result equivalent to Eq.~\eqref{eq:xiofE}, albeit with different values of $P_c$ and $L_c$.

\begin{figure}[tb]
 \includegraphics[width=\columnwidth]{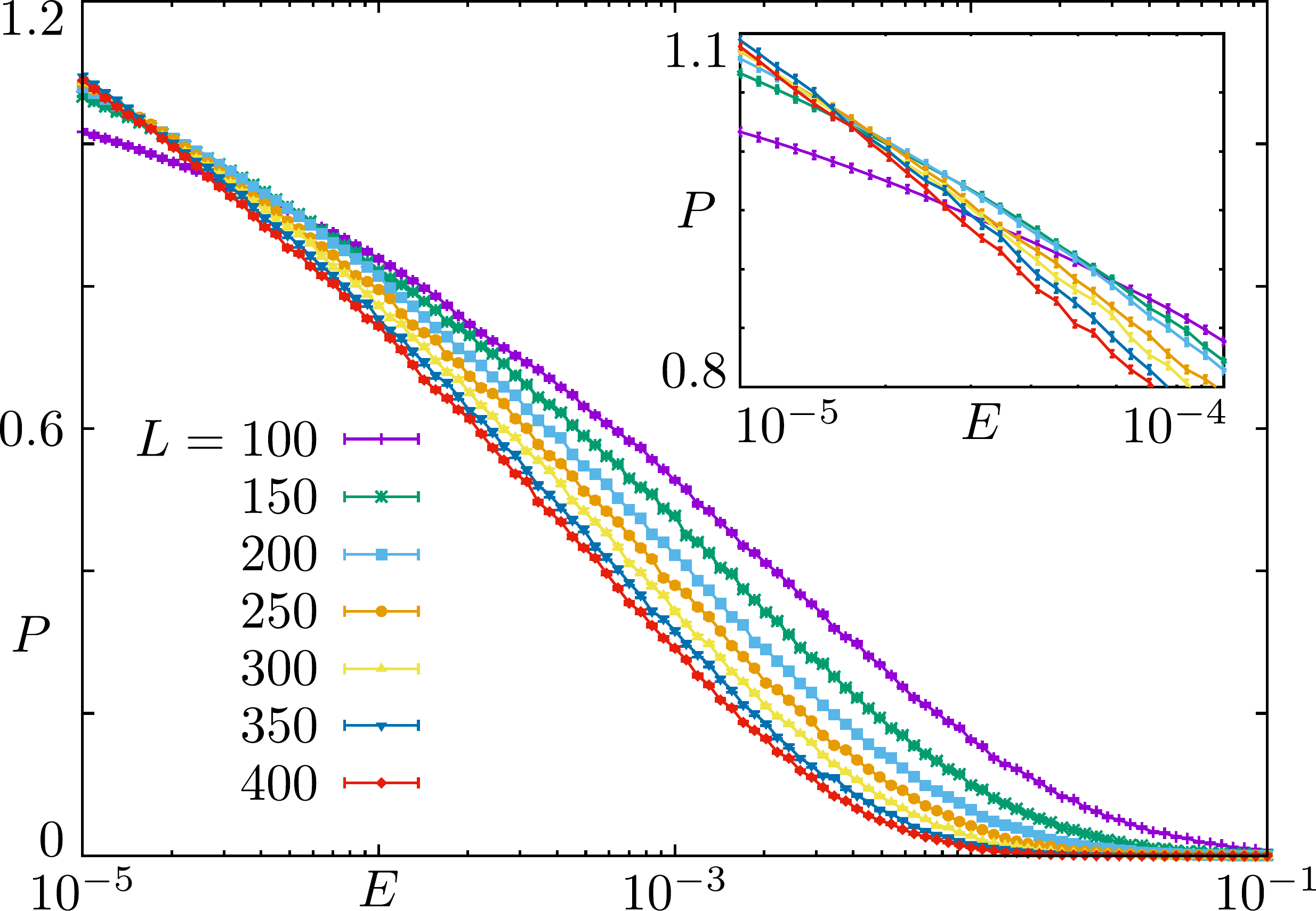}
 \caption{Average transmission probability computed at $v_2=0$, $V=1.3$, plotted as a function of system size $L$ and energy $E$ (horizontal axis, logarithmic scale). 
 Each data point was obtained using an $L\times L$ system, by averaging over $10^4$ independent disorder realizations. 
 The inset shows a closeup of the low-energy portion of the plot. 
 At small $E$, the transmission probability first increases as a function of system size, and then becomes a decreasing function of $L$ for large systems.
 \label{fig:conds_smallE}
 }
\end{figure}

\begin{figure}[tb]
 \includegraphics[width=1\columnwidth]{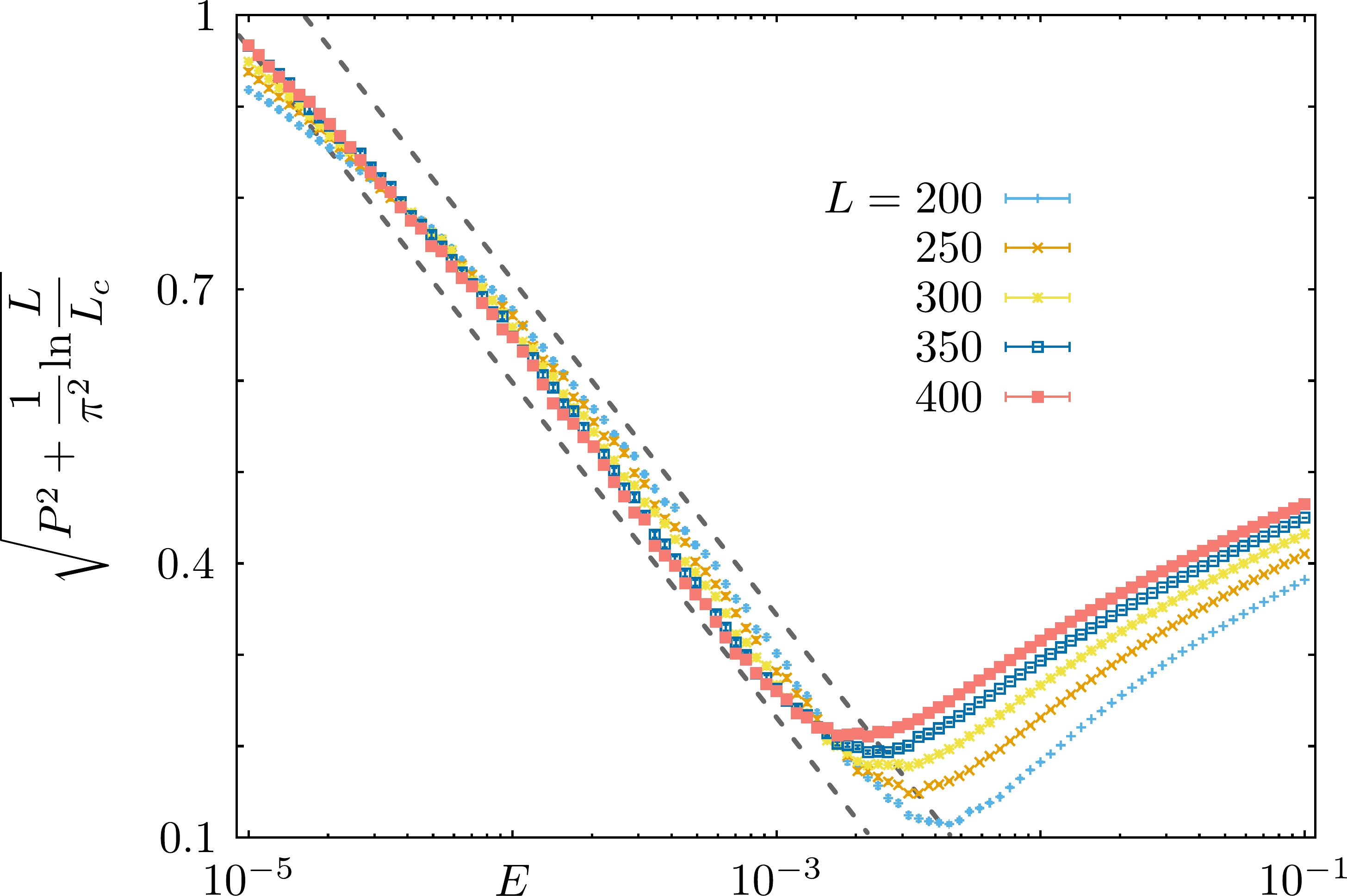}
 \caption{We rescale the data of Fig.~\ref{fig:conds_smallE} using the weak-localization result of Eq.~\eqref{eq:condA}. 
 The value of $L_c$ is given by Eq.~\eqref{eq:Lc} with $\ell_0=16$. 
 The thick dashed lines show a ${\rm ln}(1/E)$ behavior with slope equal to $1/(2\pi)$, as predicted by Eq.~\eqref{eq:gc}.\label{fig:rescaled}
 }
\end{figure}

We test the $\sigma$ model prediction by numerically determining the transmission probability $P$ at small energies and large system sizes, using a cylindrical geometry, setting as before a constant disorder strength $V=1.3$. 
Figure \ref{fig:conds_smallE} shows that, consistent with the previous discussion, the transmission probability is a nonmonotonic function of $L$ at small $E$. 
For small systems, we observe positive scaling, as expected in class D. 
For the largest system sizes, however, the scaling becomes negative, as for systems belonging to class A. 
The smaller the energy, the larger the values of $L$ required before $P$ becomes a decreasing function of system size.

We use Eq.~\eqref{eq:condA} to rescale the numerically determined transmission probabilities shown in Fig.~\ref{fig:conds_smallE}. 
This rescaling is shown in Fig.~\ref{fig:rescaled}, where we plot the energy dependence of $\sqrt{P^2+(1/\pi^2){\rm ln}(L/L_c)}$ for system sizes $L\geq200$, using $L_c$ as in Eq.~\eqref{eq:Lc}, with $\ell_0 = 16$. 
In the intermediate energy region, $10^{-4.5} \leq E \leq 10^{-3}$, the data points follow the analytical prediction of Eq.~\eqref{eq:Lc} (thick dashed lines).
Both the $\ln (1/E)$ dependence and the slope of $1/(2\pi)$ are reproduced, thus validating the $\sigma$ model results: weak antilocalization of class D and weak localization of class A. 
At the largest energies, $E\geq 10^{-2}$, we observe a sharp change of the behavior. 
This signals that the system is in the strong localization regime for these energies, i.e. transmission probability is very small and decays exponentially with system size. 
Finally, we note that for the smallest energies, $E\simeq 10^{-5}$, there is a deviation from a straight line behavior, which becomes smaller with increasing system size. 
This indicates that for the system sizes we can reach, $L \le 400$, our smallest energies $E\simeq 10^{-5}$ are close to the crossover from class D to class A, such that the weak localization behavior of Eq.~\eqref{eq:condA} has not yet fully set in.

\section{Finite temperature regimes}
\label{sec:finite_T}

As mentioned in the main text, the longitudinal conductance Eq.~\eqref{eq:cond} is an increasing function of system size in the limit of zero temperature, but shows an insulating behavior at high $T$. 
Here we study the behavior of $\kappa_{xx}$, identifying three temperature regimes, which depend on the relation between $T$ and the two crossover energy scales introduced in the main text, $E_L$ and $E_c$.

When $T<E_L$, the main contribution to the energy integral Eq.~\eqref{eq:cond} comes from states with $E\sim T$ whose diffusion length $L_E$ is larger than system size. 
These states are characterized by weak antilocalization, leading to the well known conductance scaling of a thermal metal, $\kappa_{xx}\simeq P(E=0)= {\rm const} + (1/\pi) \ln (L/\ell_0)$, as shown in Eq.~\eqref{eq:condD} and in the left panel of Fig.~\ref{fig:cond_vs_L_vary_T}.

As temperature increases, a second scaling regime occurs when $E_L < T < E_c$. 
The integral Eq.~\eqref{eq:cond} is then dominated by states with $E \sim T$ belonging to this intermediate energy range, for which the transmission probability shows weak localization, as shown in Eq.~\eqref{eq:condA}. 
We therefore find $\kappa_{xx} \approx P(E=T)  = (1/\pi) \sqrt{ (1/4) \ln^2(1/T) - \ln (T^{1/2} L/\ell_0 ) }$, where we have used Eqs.~\eqref{eq:condA} and \eqref{eq:Lc}.
The thermal conductivity decreases logarithmically with increasing $L$ or increasing $T$ in this regime, which is a consequence of weak localization. 

Finally, at temperatures $T>E_c$, the prefactor to the transmission probability in Eq.~\eqref{eq:cond} is peaked at energies for which states are strongly localized, such that $P$ is exponentially small. 
The main contribution to the energy integral then comes from $E\sim E_c(L)$, for which $P\sim1$. 
This leads to the result $\kappa_{xx}(T,L) \sim [E_c(L) / T]^2$, where $E_c \approx \exp (-2\sqrt{\ln (L/\ell_0)})$ is obtained by setting $\xi(E_c)=L$ in Eq.~\eqref{eq:xiofE}. 
In this regime, the longitudinal conductivity decreases fast with increasing temperature or increasing length. 

\bibliography{References}